\documentclass[twocolumn, amsmath,trackchanges]{aastex702}
\usepackage{bm}

\newcommand{\R}{\mathcal{R}}
\newcommand{\F}{\mathcal{F}}
\newcommand{\T}{\mathcal{T}}

\newcommand{\bb}{\bm{b}}
\newcommand{\bu}{\bm{u}}
\newcommand{\bforce}{\bm{f}}

\newcommand{\bnabla}{\bm{\nabla}}
\newcommand{\bnablar}{\bm{\nabla}_{\bm{r}}}

\newcommand{\wz}{\bm{\omega}}
\newcommand{\jz}{\bm{j}}

\newcommand{\lbracket}{\left[}
\newcommand{\rbracket}{\right]}
\newcommand{\rf}{\right)}
\newcommand{\lf}{\left(}

\begin{document}

\title{Different exact-law formulations to calculate the energy cascade rate in magnetohydrodynamic turbulence with and without mean magnetic fields}

\author[orcid=0009-0005-5156-0133,sname='Sasmal']{Ramesh Sasmal}
\affiliation{Department of Physics, Indian Institute of Technology Kanpur, Uttar Pradesh 208016, India}
\email[show]{rameshs21@iitk.ac.in}  

\author[orcid=0000-0002-3746-0989, sname='Banerjee']{Supratik Banerjee} 
\affiliation{Department of Physics, Indian Institute of Technology Kanpur, Uttar Pradesh 208016, India}
\email[show]{sbanerjee@iitk.ac.in}

%% Use the \collaboration command to identify collaborations. This command
%% takes an optional argument that is either a number or the word "all"
%% which tells the compiler how many of the authors above the command to
%% show. For example "\collaboration[all]{(DELVE Collaboration)}" wil include
%% all the authors above this command.
%%
%% Mark off the abstract in the ``abstract'' environment. 
\begin{abstract}
At scales much larger than the ion inertial length and ion-gyroscale, space-plasma turbulence is often described within the framework of incompressible magnetohydrodynamics (MHD). Using high-resolution direct numerical simulations with $1024^3$ grid points, we perform the first systematic numerical comparison of three exact-law formulations for the total energy cascade in homogeneous, but not necessarily isotropic, MHD turbulence: the Yaglom–Monin (YM), Banerjee–Galtier (BG), and two-point correlator formulations. Despite directional variations, the angle-averaged energy transfer rates for all three formulations show excellent agreement for zero, moderate, and strong background magnetic fields. These results clearly suggest the usefulness of alternative BG and correlator formulations for estimating the energy cascade rates in turbulent space plasmas when the required spatial derivatives are accessible through multi-spacecraft measurements. We also argue that the isotropic Politano-Pouquet (PP) formulation does not provide a true directional transfer rate and therefore its application as a single-direction energy transfer rate estimator leads to inaccurate values, especially under anisotropic conditions. Decomposing the exact formulations into RMHD and non-RMHD contributions, we also show that the RMHD cascade rates cannot be obtained merely by increasing the strength of the mean magnetic field in incompressible magnetohydrodynamic turbulence. Finally, a term-by-term analysis of different formulations reveals how different nonlinear interaction channels coexist and collaborate with each other to maintain the same energy cascade rate across the inertial scales.  
\end{abstract}

%% Keywords should appear after the \end{abstract} command. 
%% The AAS Journals now uses Unified Astronomy Thesaurus (UAT) concepts:
%% https://astrothesaurus.org
%% You will be asked to selected these concepts during the submission process
%% but this old "keyword" functionality is maintained in case authors want
%% to include these concepts in their preprints.
%%
%% You can use the \uat command to link your UAT concepts back its source.

%%\keywords{\uat{Turbulence} --- \uat{Magnetohydrodynamics} --- \uat{Exact relation}}

\section{Introduction}

The multiscale complexity due to the highly nonlinear nature of a turbulent flow is an undisputed truth. However, the study of fully developed turbulence becomes particularly interesting when it can be characterized in terms of a universal diagnostic. Far from the macroscopic length scales which are affected by the geometry of the flow and the nature of the forcing, and the small scales where the viscous effects dominate, a turbulent flow is characterized by a nonlinear cascade of energy from one scale to its subsequent smaller scale with constant transfer rate \citep{Frisch_1995, Pope_2000, banerjee_thesis}. 

For homogeneous, isotropic, and stationary turbulence, this transfer rate is equal to the average energy injection rate (or the average energy dissipation rate) $\varepsilon$ and  is analytically calculated from the so-called exact relations, where the slope of the linear scaling of third-order structure functions with length scale $r$ is proportional to $\varepsilon$.  In hydrodynamic (HD) turbulence, using the tensorial formalism proposed by Von Karman-Howarth \citep{karman_1938}, Kolmogorov derived the $4/5$ law given by  $\varepsilon=-(5S_{rrr}^u)/(4r)$, where $S_{rrr}^u= \langle(\delta u_{r})^3\rangle$ denotes the third-order longitudinal velocity structure function with $\delta\bu = \bu(\bm{x}+\bm{r})-\bu(\bm{x})$, and $\delta u_{r}=\delta\bu\cdot\hat{r}$, $\hat{r}$ being the unit vector along the separation vector $\bm{r}$ \citep{kolmogorov_1941}. However, in the case of homogeneous and isotropic magnetohydrodynamic (MHD) turbulence, it is not straightforward to express $\varepsilon$ purely in terms of third-order longitudinal structure functions. Relaxing the assumption of isotropy, one may follow the Yaglom-Monin (hereafter YM) formulation to express the cascade rate in divergence form as $\varepsilon=\bnabla_{\bm{r}}\cdot\bm{\mathcal{F}}$ with $\bm{\mathcal{F}}= -\langle(\delta \bu)^2\delta \bu\rangle/4$ and $- \left\langle\lbracket\lf\delta \bu\rf^2+(\delta\bb)^2\rbracket\delta\bu-2\lf\delta\bu\cdot\delta\bb\rf\delta\bb\right\rangle/4$ for HD and MHD, respectively \citep{Antonia_1997, politano_1998b, podesta_2008}. Further addition of isotropy leads to  $\varepsilon=(3/r) \mathcal{F}_r$, where $\mathcal{F}_r = \bm{\mathcal{F}}\cdot\hat{r}$, which is popularly known as Politano-Pouquet (hereafter PP) formulation \citep{politano_1998a}. In the current paper, we designate the Yaglom-Monin-like divergence form as YM, whereas the isotropic form will be labeled as the PP form. Beyond ordinary HD and MHD turbulence, the YM formulation has been used to derive exact laws for the energy transfer in many complex fluid models such as reduced MHD (RMHD) \citep{sasmal_2025}, finite-Larmor-radius MHD \citep{sasmal_2026}, Hall MHD \citep{ferrand_2019}, electron MHD \citep{celani_1998a, celani_1998b}, stratified turbulence \citep{augier_2012}, binary fluids \citep{pan_2022}, \textit{etc}. Despite the considerable popularity of the YM formulation, the cascade rates in several complex fluid systems, such as ferrofluids, compressible HD and MHD, compressible Hall MHD, \textit{etc.}, cannot be expressed purely in the divergence form \citep{banerjee_2011, banerjee_2013, banerjee_2014}. 

It was only about a decade ago that an alternative formulation for calculating $\varepsilon$ was proposed, in which, instead of divergence, the cascade rate $\varepsilon$ is expressed in terms of mixed second-order structure functions (hereafter BG).  According to this formulation, cascade rates $\varepsilon$ for HD and MHD are obtained as $\varepsilon=\left\langle\delta\bu\cdot\delta\lf\bu\times\wz\rf\right\rangle/2$ and $\varepsilon=\left\langle\delta\bu\cdot\delta\lf\bu\times\wz+\jz\times\bb\rf+\delta\jz\cdot\delta\lf\bu\times\bb\rf\right\rangle/2$, respectively.  Unlike the YM formulation, where the cascade rate is expressed as a divergence of a flux, the BG formulation recognizes the cascade rate as a measure of departure from the Beltrami and dynamically aligned states \citep{banerjee_2017, banerjee_2023}.  In addition to fluid flows, where the YM form can be obtained \citep{pan_2022, halder_2025}, the BG formulation can also be extended to the case of ferrofluids, compressible HD and MHD flows \citep{banerjee_2018, mouraya_2019}. A third form can be obtained directly in terms of the two-point correlators as shown in Sec. \ref{exact}. 

The exact laws are extremely important tools to determine the turbulent cascade rate in natural systems such as neutral fluids, space plasmas, fusion plasmas, \textit{etc}. In space plasmas, the knowledge of $\varepsilon$ is particularly important as it provides an estimate of the energy transfer rate available for heating in those plasma environments. However, many important solar missions use a single spacecraft where the \textit{in-situ}
plasma and magnetic field data are often provided in the form of a time series along the spacecraft trajectory. Using conventional techniques, it is therefore impossible to calculate the three-dimensional differential operator $\bnablar$ of the YM form to determine $\varepsilon$ from single-spacecraft data. For simplicity, one can neglect the angular dependence of $\bm{\mathcal{F}}$ and determine $\varepsilon$ from the isotropic PP form \citep{valvo_2007, MacBride_2008, Coburn_2014}. If the mean plasma velocity is much larger than both the Alfv\'en speed and the speed of the spacecraft, one can use Taylor's hypothesis, where time lags of the corresponding time series can be used as a reasonable proxy for the length scales. However, in the presence of a non-negligible mean magnetic field \citep{matthaeus_1996, horbury_2008, deepali_2021, sakshee_2022}, the angular dependence of $\bm{\mathcal{F}}$ cannot be ignored and the estimation of $\varepsilon$ from the PP formulation may be questionable. This problem persists even if one uses multi-spacecraft data. However, in that case, an indirect way to measure the cascade rate using the YM formulation is to assume axisymmetry around the mean magnetic field as described in \cite{osman_2011}. Despite being an interesting effort, the assumption of axisymmetry may not be valid in the solar wind plasma, where the mean magnetic field is only moderately strong. In addition, the multi-spacecraft measurements provide limited polar angle coverage of $0^\circ-10^\circ$ and $80^\circ-90^\circ$, which can affect the reliability of the estimated cascade rate. Recently, a method of lag polyhedral derivative ensemble (LPDE) has been proposed by \cite{Pecora_2023} to estimate the energy cascade rate in the YM formulation using lags between multi-spacecraft. However, spacecraft lags cover a limited range of length scales roughly $5$ to $25$ $km$ for the MMS spacecraft, which does not correspond to the inertial range energy cascade rate in space plasmas at $1$ au. As an interesting and simple solution for the aforementioned problem, one may consider the BG or the correlator formulation to calculate $\varepsilon$. However, the BG formulation involves local curls of $\bu$ and $\bb$, whereas the correlator formulation requires spatial gradients of $\bu$ and $\bb$. Thus, neither BG nor the correlator formulation can be used to determine $\varepsilon$ using single-spacecraft data. This obstacle can still be overcome if we can calculate the spatial derivatives using
multi-spacecraft techniques \citep{dunlop_1988, dunlop_2015, singh_2026}.

Despite a plethora of numerical and observational studies regarding the YM and PP formulations, no systematic numerical study has been performed involving BG and correlator formulations in MHD turbulence and comparing them with YM and PP formulations to examine their robustness with and without the presence of a mean magnetic field. Note that the work by \cite{ferrand_2019}
performed a comparative study just for the Hall term and not for the entire exact law. The current study is probably the first systematic DNS comparison of the aforementioned formulations to investigate their equivalence as a function of mean magnetic field strengths. Unlike these three formulations, the PP formulation shows a non-trivial angular dependence which increases with the strength of the mean magnetic field. The main objective of this paper is to show the fact that instead of the PP formulation, either the BG or the correlator formulations, wherever feasible, may give a reliable estimate of incompressible energy cascade rates in MHD turbulence with or without a  mean magnetic field. The validation of the same can serve as a benchmark for the applicability of the BG and the correlator formulations to estimate the turbulent heating rate in space and astrophysical plasmas. 

The paper is organized as follows: in Sec. \ref{governing}, we introduce the governing equations of incompressible MHD. Sec. \ref{exact} presents different formulations of exact relations for the energy cascade rates. The numerical details are given in Sec. \ref{numerics}. In Sec. \ref{result}, we compare the transfer rates calculated from YM, BG, correlator, and PP formulations, study their angular dependence for different values of the mean magnetic field, check the validity of RMHD approximations within the incompressible MHD framework, and present a term-by-term analysis of each formulation. Finally, in Sec. \ref{conclusion}, we summarize our results and conclude. 

\begin{figure*}
    \centering
    \includegraphics[width=0.9
    \textwidth]{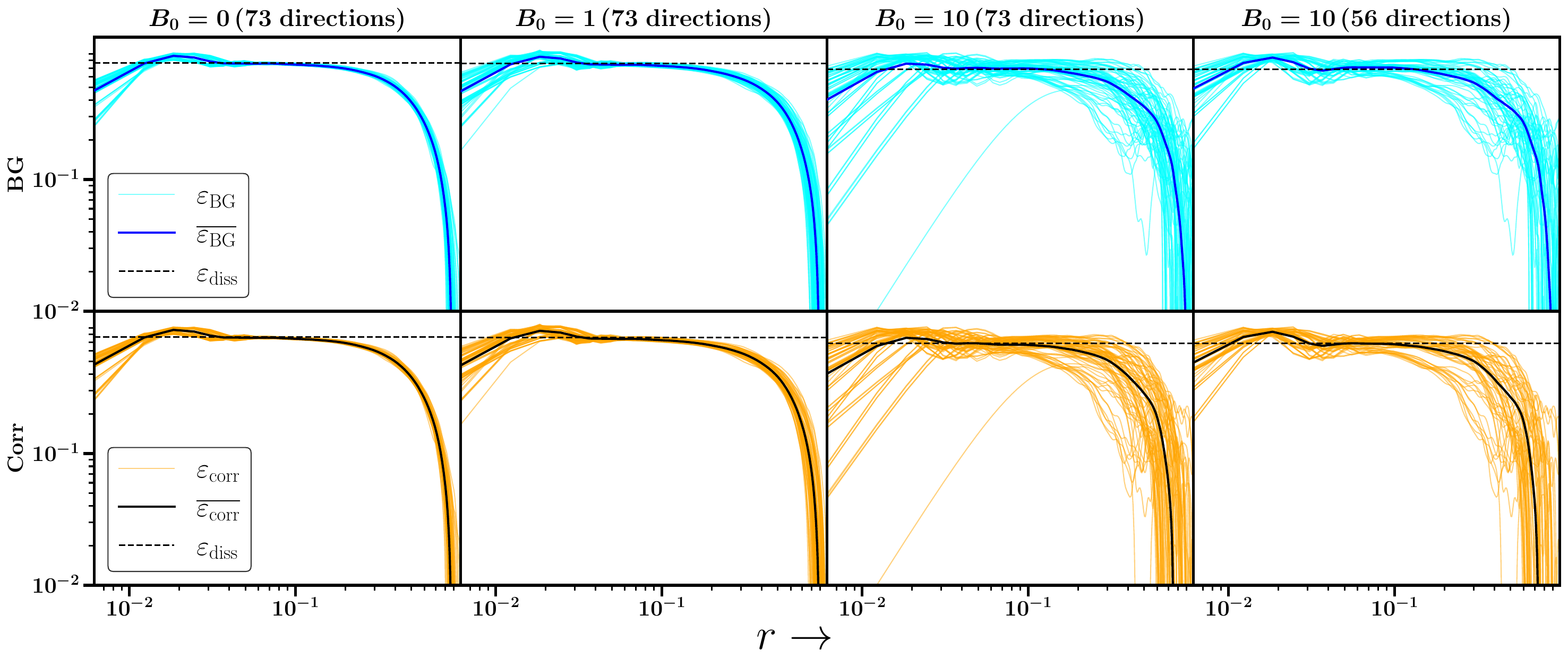}
    \caption{Directional dependence of the energy cascade rates in the BG and Corr formulations. Top panels: Energy cascade rates along individual directions, \(\varepsilon_{\mathrm{BG}}\) (aqua lines), and their directional averages, \(\overline{\varepsilon_{\mathrm{BG}}}\) (blue line), obtained using the BG formulation for different strengths of \(B_0\). Bottom panels: Corresponding energy cascade rates obtained using the Corr formulation, \(\varepsilon_{\mathrm{corr}}\) (orange lines), and their directional averages, \(\overline{\varepsilon_{\mathrm{corr}}}\) (black line). For \(B_0=10\), the fourth column additionally shows the cascade rates along 56 directions ($x, \, y, \, z$) satisfying \(x^2+y^2\geq z^2\), out of the 73 directions considered in the other panels.}
    \label{directions}
\end{figure*}

\section{Governing equations}\label{governing}
In the presence of a uniform background mean magnetic field, the governing equations of incompressible MHD are given by
\small{
\begin{align}
    &\bnabla\cdot\bu = 0,\\
    &\bnabla\cdot\bb = 0,\\
    \frac{\partial \bu}{\partial t}+(\bu\cdot\bnabla)\bu=&-\bnabla p*+ [\left(\bm{B_0}+\bb\right)\cdot\bnabla]\bb+\bm{f}\nonumber\\
    &\hspace{1cm}+\nu(-1)^{\alpha+1}\bnabla^{2\alpha}\bu,\text{ and }\\
    \frac{\partial \bb}{\partial t}+(\bu\cdot\bnabla)\bb =& [\left(\bm{B_0}+\bb\right)\cdot\bnabla]\bu +\eta(-1)^{\alpha+1}\bnabla^{2\alpha}\bb,
\end{align}}

\noindent where $\bu$ represents the fluctuating velocity field, $\bm{B_0}$ and $\bb$ denote the background and fluctuating magnetic fields in Alfv\'en units, respectively,  $p*=p+(\bm{B_0}+\bb)^2/2$, where $p$ is the fluid pressure, $\bm{f}$ represents a forcing term, $\nu$ and $\eta$ are the hyper-viscosity and the hyper-resistivity coefficients, respectively, with $\alpha$ denoting the corresponding hyperdiffusive order. Note that the case $\alpha=1$ corresponds to the standard fluid dissipation. In the unforced, inviscid limit, incompressible MHD conserves two invariants: (i) the total energy, $\int \left(\bu^2+\bb^2\right)/2 \,d\tau$ and (ii) the cross helicity, $\int (\bu\cdot\bb)\,d\tau$. \footnote{In the presence of a uniform mean magnetic field, the fluctuating magnetic helicity $\int (\bm{a}\cdot\bb)\,d\tau$ is not an inviscid invariant of the system. Here, $\bm{a}$ denotes the vector potential.}
In the present study, we only focus on the total energy and its universal cascade within the inertial range of length scales.

\section{Different formulations of the exact law}\label{exact}
To determine the total energy cascade rate $\varepsilon$ from the governing equations, we construct the symmetric two-point correlator for the total energy as
\begin{equation}
    \R_E = \frac{1}{2}\left\langle\bu\cdot\bu'+\bb\cdot\bb'\right\rangle,
\end{equation}
\noindent where primed and unprimed quantities represent the corresponding field variables at positions $\bm{x}+\bm{r}$ and $\bm{x}$, respectively. By straightforward algebra, under the assumption of statistical homogeneity, one obtains \citep{podesta_2008}
\begin{align}
    \partial_t \R_E
    =&\frac{1}{4}\bnablar \cdot\left\langle\lbracket\lf\delta \bu\rf^2+(\delta\bb)^2\rbracket\delta\bu-2\lf\delta\bu\cdot\delta\bb\rf\delta\bb\right\rangle\nonumber\\
    &+ F_E + D_E,\label{energy1}
\end{align}
\noindent where $\left\langle(\cdot)\right\rangle$ represents the ensemble average, $F_E=\left\langle\bu\cdot\bforce'+\bu'\cdot\bforce\right\rangle/2$ the average energy injection rate, $D_E = \lf D_u + D_m\rf/2$  the average contribution due to small-scale dissipation with  $D_u = (-1)^{\alpha+1}\left\langle\nu\lbracket\lf\bu\cdot\bnabla'^{2\alpha}\bu'\rf+\lf\bu'\cdot\bnabla^{2\alpha}\bu\rf\rbracket\right\rangle$ and $D_m =(-1)^{\alpha+1} \left\langle\eta\lbracket\lf\bb\cdot\bnabla'^{2\alpha}\bb'\rf+\lf\bb'\cdot\bnabla^{2\alpha}\bb\rf\rbracket\right\rangle$, and for a variable $X$, $\delta X = X'-X$. For statistically stationary turbulence, the left-hand side of Eq. (\ref{energy1}) vanishes. Moreover, far from the small dissipation scales, in the limit of infinite Reynolds number, one can neglect $D_E$ within the inertial range of length scales. Assuming $\bm{f}$ to be effective at large scales, one obtains  $F_E=\left\langle\bu\cdot\bforce'+\bu'\cdot\bforce\right\rangle/2\approx\left\langle\bu\cdot\bm{f}\right\rangle=\varepsilon_\mathrm{inj}$
in the inertial range, where $\varepsilon_{\mathrm{inj}}$ denotes the average one-point energy injection rate. Taking all these assumptions in the aforementioned order, Eq. (\ref{energy1}) finally reduces to 
\begin{equation}
    \varepsilon_{\mathrm{inj}} = -\frac{1}{4}\bnablar \cdot\left\langle\lbracket\lf\delta \bu\rf^2+(\delta\bb)^2\rbracket\delta\bu-2\lf\delta\bu\cdot\delta\bb\rf\delta\bb\right\rangle\label{energy2}.
\end{equation}
\noindent Note that the left-hand side of Eq. (\ref{energy2}) can be replaced by the average one-point energy dissipation rate, $\varepsilon_\mathrm{diss}$, due to the balance between average energy injection and dissipation in a statistically stationary state. The right-hand side of Eq. (\ref{energy2}) can also be written in terms of the two-point correlator as 
\begin{equation}
   \varepsilon_{\mathrm{inj}}= - \lbracket\T_1(\bm{r})+\T_2(\bm{r}) + \T_3(\bm{r})+\T_4(\bm{r})\rbracket,\label{energy3}
\end{equation}
\noindent where 

\begin{align}
    \T_1&=-\left\langle\bu'\cdot\lf\bu\cdot\bnabla\rf\bu+\bu\cdot\lf\bu'\cdot\bnabla'\rf\bu'\right\rangle/2,\\
\T_2&=\left\langle\bu'\cdot\lf\bb\cdot\bnabla\rf\bb+\bu\cdot\lf\bb'\cdot\bnabla'\rf\bb'\right\rangle/2,\\
\T_3&=-\left\langle\bb'\cdot\lf\bu\cdot\bnabla\rf\bb+\bb\cdot\lf\bu'\cdot\bnabla'\rf\bb'\right\rangle/2,\text{ and}\\
\T_4&=\left\langle\bb'\cdot\lf\bb\cdot\bnabla\rf\bu+\bb\cdot\lf\bb'\cdot\bnabla'\rf\bu'\right\rangle/2.
\end{align}
\noindent A third alternative form of Eq. (\ref{energy2}) is given as \citep{banerjee_2017}
\begin{align}
    \varepsilon_{\mathrm{inj}}=\frac{1}{2}\left\langle\delta\bu\cdot\delta\lf\bu\times\wz+\jz\times\bb\rf+\delta\jz\cdot\delta\lf\bu\times\bb\rf\right\rangle,\label{energy4}
\end{align}
\noindent where $\wz = \bnabla\times \bu$ and $\jz=\bnabla\times\bb$. Under the assumptions of statistical homogeneity and stationarity we have $\varepsilon_{\mathrm{diss}}=\varepsilon_{\mathrm{inj}}\equiv\varepsilon$ and one obtains three expressions for the cascade rate of the total energy as (i) the YM form: $\varepsilon=\varepsilon_{\mathrm{YM}}$, (ii) the correlator form (Corr hereinafter): $\varepsilon=\varepsilon_{\mathrm{corr}}$, and (iii) the BG form: $\varepsilon = \varepsilon_{\mathrm{BG}}$, where 
\begin{align}
    &\varepsilon_{\mathrm{YM}}=-\frac{1}{4}\bnablar \cdot\left\langle\lbracket\lf\delta \bu\rf^2+(\delta\bb)^2\rbracket\delta\bu-2\lf\delta\bu\cdot\delta\bb\rf\delta\bb\right\rangle,\label{ym}\\
    &\varepsilon_{\mathrm{corr}}=\frac{1}{2}\left\langle\lbracket\bu'\cdot\lf\bu\cdot\bnabla\rf\bu+\bu\cdot\lf\bu'\cdot\bnabla'\rf\bu'\rbracket\right.\nonumber\\
    &\hspace{.7cm}-\left.\lbracket\bu'\cdot\lf\bb\cdot\bnabla\rf\bb+\bu\cdot\lf\bb'\cdot\bnabla'\rf\bb'\rbracket\right.\nonumber\\
&\hspace{.7cm}\left.+\lbracket\bb'\cdot\lf\bu\cdot\bnabla\rf\bb+\bb\cdot\lf\bu'\cdot\bnabla'\rf\bb'\rbracket\right.\nonumber\\
&\hspace{.7cm}\left.-\lbracket\bb'\cdot\lf\bb\cdot\bnabla\rf\bu+\bb\cdot\lf\bb'\cdot\bnabla'\rf\bu'\rbracket\right\rangle, \text{ and} \label{corr}\\
&\varepsilon_{\mathrm{BG}}=\frac{1}{2}\left\langle\delta\bu\cdot\delta\lf\bu\times\wz+\jz\times\bb\rf+\delta\jz\cdot\delta\lf\bu\times\bb\rf\right\rangle.\label{bg}
\end{align}
\noindent Note that the presence of $\bm{B_0}$ does not modify the above expressions for the cascade rate of the total energy. Moreover, under the assumption of isotropy, integrating Eq. (\ref{energy2}) over a sphere of radius $r$, one obtains the PP form for the energy cascade as \citep{politano_1998a}
\begin{equation}
    \varepsilon_{\mathrm{PP}}= -\frac{3}{4r}\left\langle\lbracket\lf\delta \bu\rf^2+(\delta\bb)^2\rbracket\delta u_r-2\lf\delta\bu\cdot\delta\bb\rf\delta b_r\right\rangle, \label{politano}
\end{equation}
\noindent where $\delta u_r= \delta\bu\cdot\hat{r}$ and $\delta b_r= \delta\bb\cdot\hat{r}$.
\begin{figure*}
    \centering
    \includegraphics[width=\textwidth]{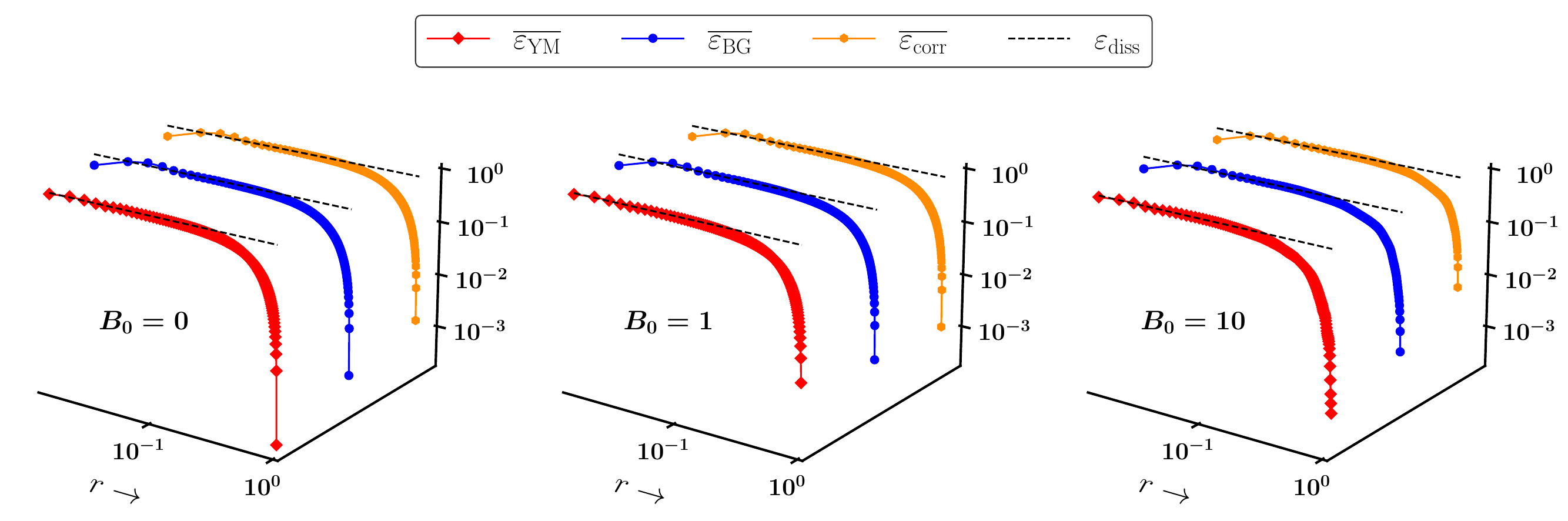}
    \caption{Comparison of energy cascade rates obtained from different exact law formulations. Three-dimensional projections of the average energy cascade rates obtained using the YM, BG, and Corr formulations, $\overline{\varepsilon_{\mathrm{YM}}}$ (red line), $\overline{\varepsilon_{\mathrm{BG}}}$ (blue line), and $\overline{\varepsilon_{\mathrm{corr}}}$ (orange line), respectively, for the mean magnetic field strengths of $B_0=0,\,1,$ and $10$. The black dashed line indicates the corresponding average one-point dissipation rate, $\varepsilon_{\mathrm{diss}}$ and $r$ represents the length scales.}
    \label{3d_comparison}
\end{figure*}
\section{Numerical details}\label{numerics}
In the following, we use data of forced MHD turbulence with uniform mean magnetic field $\bm{B_0}=B_0\hat{z}$ where $B_0=0,\, 1\,(B_0/b_\mathrm{rms}=1.2), \, \text{and} \,10\, (B_0/b_\mathrm{rms}=12.7)$. The datasets are downloaded from the Smart-Turb turbulent database \citep{smart_turb}. The data were generated by DNS of 3D incompressible MHD equations in a cubic, triply periodic domain of size $L_\mathrm{box}=2\pi$ with $1024^3$ grid points and hyperviscous and hyperresistive dissipations with $\alpha=5$. For our analysis, we use data  with $\nu=\eta=5\times10^{-23}$. Each MHD turbulence simulation was sustained by Ornstein-Uhlenbeck forcing, with minimal injection of cross helicity, active in the wavenumber band $k\in [2.5, 5.0]$. Turbulence simulations with $B_0=0$ and $1$ attain a statistically stationary state, whereas the simulation with $B_0=10$ exhibits a linear growth in kinetic energy. However, the temporal evolution of the kinetic energy spectrum indicates that the non-stationarity is primarily associated with the low-wavenumber modes, where energy continues to accumulate at large scales but small-scale dissipation shows a stationary behavior over time. In this regard, the simulation with $B_0=10$ can be assumed to have attained a quasi-stationary state. The snapshots that we have downloaded have kinetic ($\varepsilon_u$) and magnetic ($\varepsilon_b$) energy dissipation rates of ($\varepsilon_u, \,\varepsilon_b$) = ($0.32,\,0.45$), ($0.33, \,0.44$), and ($0.30,\,0.38$)  for the simulations with $B_0=0,\, 1$ and $10$, respectively. More details about the simulations can be found in \cite{capocci_2025}.

\section{Results and discussion}\label{result}
\subsection{Comparison of different exact-law formulations: directional dependence of the cascade rates}
In this section, we compare different exact relations given by Eqs. (\ref{ym})–(\ref{bg}) and investigate whether they consistently describe a universal inertial-range energy cascade for $B_0=0$, $1$, and $10$. We first investigate the nonlinear transfer rates obtained from the BG and Corr formulations for varying strengths of $B_0$. From Eqs. (\ref{corr}) and (\ref{bg}), it follows that these rates are, in general, functions of the separation vector $\bm{r}$ and hence, depend on the spherical coordinates $r$, $\theta$ (the co-latitude), and $\phi$ (the azimuthal angle). In order to verify the existence of a Kolmogorov-like energy cascade, we first seek to obtain a direction-independent cascade rate that depends only on the separation scale $r$. To eliminate the angular dependence of the cascade rates obtained from the BG and Corr formulations, one uses angle averaging given by 
\begin{align}
    \overline{\varepsilon_\mathrm{BG}}(r)&= \langle\varepsilon_\mathrm{BG}\rangle_{\theta, \, \phi}= \frac{1}{4\pi}\int^{2\pi}_0\int^\pi_0\varepsilon_\mathrm{BG}\sin \theta\, d\theta d\phi,\\
    \overline{\varepsilon_\mathrm{corr}}(r)&= \langle\varepsilon_\mathrm{corr}\rangle_{\theta, \, \phi}= \frac{1}{4\pi}\int^{2\pi}_0\int^\pi_0\varepsilon_\mathrm{corr}\sin \theta\, d\theta d\phi.
\end{align}
\begin{figure*}
    \centering
    \includegraphics[width=0.9
    \textwidth]{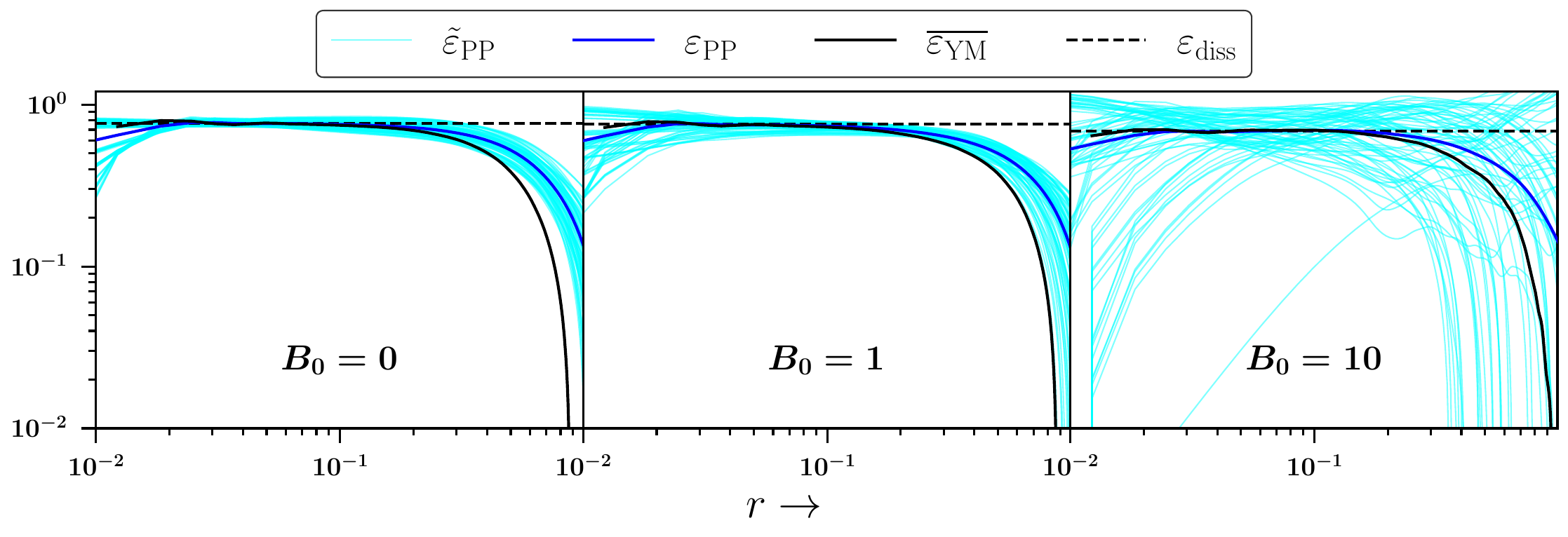}
    \caption{Angular dependence of pseudo-PP estimates along individual directions, $\tilde{\varepsilon}_\mathrm{PP}$ (aqua lines) for increasing strengths of mean magnetic field and comparison between $\varepsilon_\mathrm{PP}$ (blue line) with $\overline{\varepsilon_\mathrm{YM}}$ (black line).}
    \label{pp_vs_ym}
\end{figure*}
\noindent However, in practice, approximate estimates of direction-independent cascade rates are obtained by employing an angle-averaging method originally proposed by \citet{taylor_2003} and subsequently used in previous studies \citep{ferrand_2021, pan_2025}. In particular, $\varepsilon_\mathrm{BG}$ and $\varepsilon_{\mathrm{corr}}$ are evaluated along approximately uniformly distributed 73 directions spanned by the vectors $(1, 0, 0),\, (1, 1, 0), \, (1, 1, 1), \, (2, 1, 0),\, (2, 1, 1),\, (2, 2, 1)$, $ (3, 1, 0),\, (3, 1, 1)$ and their permutations of the sign and index of their coordinates, while the opposite vectors are discarded. The angle-averaged transfer rates are finally obtained by $\overline{\varepsilon_\mathrm{BG}}\approx \sum\varepsilon_\mathrm{BG}/73 \text{ and }
    \overline{\varepsilon_\mathrm{corr}}\approx \sum\varepsilon_\mathrm{corr}/73$. In Fig. (\ref{directions}), we plot $\varepsilon_\mathrm{BG}$ and $\varepsilon_{\mathrm{corr}}$ evaluated along the 73 directions mentioned above with their average cascade rates $\overline{\varepsilon_\mathrm{BG}}$ and $\overline{\varepsilon_\mathrm{corr}}$, respectively. Since the maximum useful scale is half the simulation domain due to periodicity, and in the presence of a strong mean magnetic field, the dominant cascading modes are present in the plane perpendicular to it, for the case $B_0=10$, we separately show the cascade rates for $56$ directions ($x, y, z$) out of the $73$ directions that satisfy $x^2+y^2\geq z^2$ as described in \cite{ferrand_2021}. The directional transfer rates $\varepsilon_\mathrm{BG}$ and $\varepsilon_\mathrm{corr} $ for $B_0=0$ are found to be nearly identical for all the 73 directions which can be expected for a well-separated inertial-range energy cascade. For $B_0=1$, a very weak spread is found in the inertial range, whereas for $B_0=10$, the energy cascade rates exhibit a clear spread including an entirely disparate energy transfer rate along $\bm{B_0}$. The spread becomes less pronounced when we measure them along the $56$ directions mentioned above. This behavior may be attributed to the mean-field induced anisotropy which may reduce the effective separation between the energy containing and the inertial scales, particularly along the mean magnetic field, and therefore the dependence of  $\varepsilon_\mathrm{BG}$ and $\varepsilon_\mathrm{corr}$ on $\theta$ and $\phi$ cannot be neglected. Upon plotting the angle-averaged transfer rates $\overline{\varepsilon_\mathrm{BG}}$ and $\overline{\varepsilon_\mathrm{corr}}$ in 
    Fig. (\ref{directions}), we observe the existence of a range of scales across which both remain fairly constant irrespective of the strength of $B_0$. In addition, both of them match remarkably with the average one-point dissipation rate $\varepsilon_\mathrm{diss}$ over that range of scales. This confirms that angle-averaged BG and Corr formulations sustain stationary Kolmogorov-like energy cascade with constant cascade rate across the inertial scales. For the run with $B_0=10$, despite the absence of global stationarity, the angle-averaged nonlinear energy transfer rate across the inertial scales shows a close agreement with $\varepsilon_\mathrm{diss}$, which is consistent with a quasi-stationary energy cascade. However, with the increase of numerical resolution, we can, in principle, attain a turbulent flow with higher Reynolds numbers and may obtain a less contaminated inertial range where all the directional transfer rates would ideally collapse to their angle-averaged values irrespective of the strength of $B_0$.

We now include $\varepsilon_\mathrm{YM}(\bm{r})$ in our investigation, which is given by
\begin{equation}
  \varepsilon_\mathrm{YM}=\frac{1}{r^2}\frac{\partial}{\partial r}(r^2 F_r) + \frac{1}{r \sin \theta}\left[\frac{\partial}{\partial \theta}(\sin \theta\, F_\theta)+\frac{\partial \, F_\phi}{\partial \phi}\right],\label{divergence}
\end{equation}
\noindent where $\bm{F}(\bm{r})=-(1/4)\left\langle[\lf\delta \bu\rf^2+(\delta\bb)^2]\delta\bu-2\lf\delta\bu\cdot\delta\bb\rf\delta\bb\right\rangle$, $F_r=\bm{F}\cdot\hat{r}$, $F_\theta=\bm{F}\cdot\hat{\theta}$, and $F_\phi=\bm{F}\cdot\hat{\phi}$. In order to verify whether the YM formulation supports Kolmogorov-like energy cascade, one requires to calculate $\overline{\varepsilon_\mathrm{YM}} (r)$. However, unlike the BG and Corr formulations, it is cumbersome to directly calculate the derivatives with respect to $ \theta$ and $\phi$. Now, the angle-averaged cascade rate is given by 
\begin{align}
&\overline{\varepsilon_\mathrm{YM}}=\langle\varepsilon_\mathrm{YM}\rangle_{\theta, \, \phi}\nonumber\\
    &=\left\langle\frac{1}{r^2}\frac{\partial}{\partial r}(r^2 F_r) + \frac{1}{r \sin \theta}\left[\frac{\partial}{\partial \theta}(\sin \theta\, F_\theta)+\frac{\partial \, F_\phi}{\partial \phi}\right]\right\rangle_{\theta,\, \phi}
\end{align}
\noindent and furthermore, one can show that
\begin{align}
    &\int^{2\pi}_0\int^\pi_0\frac{1}{ \sin \theta}\frac{\partial}{\partial \theta}(\sin \theta\, F_\theta)\sin \theta d\theta d\phi= 0, \text{and}\label{theta}\\
    &\int^{2\pi}_0\int^\pi_0\frac{1}{ \sin \theta}\frac{\partial \, F_\phi}{\partial \phi}\sin \theta d\theta d\phi=0.\label{phi}
\end{align}

\noindent Using these relations, one finally obtains $\overline{\varepsilon_\mathrm{YM}}=(1/r^2)(\partial/\partial r)r^2 \langle F_r\rangle_{\theta, \, \phi}$. In Fig. (\ref{3d_comparison}), for the cases with $B_0=0$, $1$, and $10$, we plot the angle-averaged cascade rates obtained from the different exact law formulations along with $\varepsilon_\mathrm{diss}$. The results exhibit an excellent agreement among all the formulations for different $B_0$. This is the main result of this paper. It clearly states that any of these three formulations can be equivalently used to obtain the angle-averaged cascade rate in incompressible MHD turbulence irrespective of the strength of $B_0$.
\begin{figure*}
    \centering
    \includegraphics[width=0.8
    \textwidth]{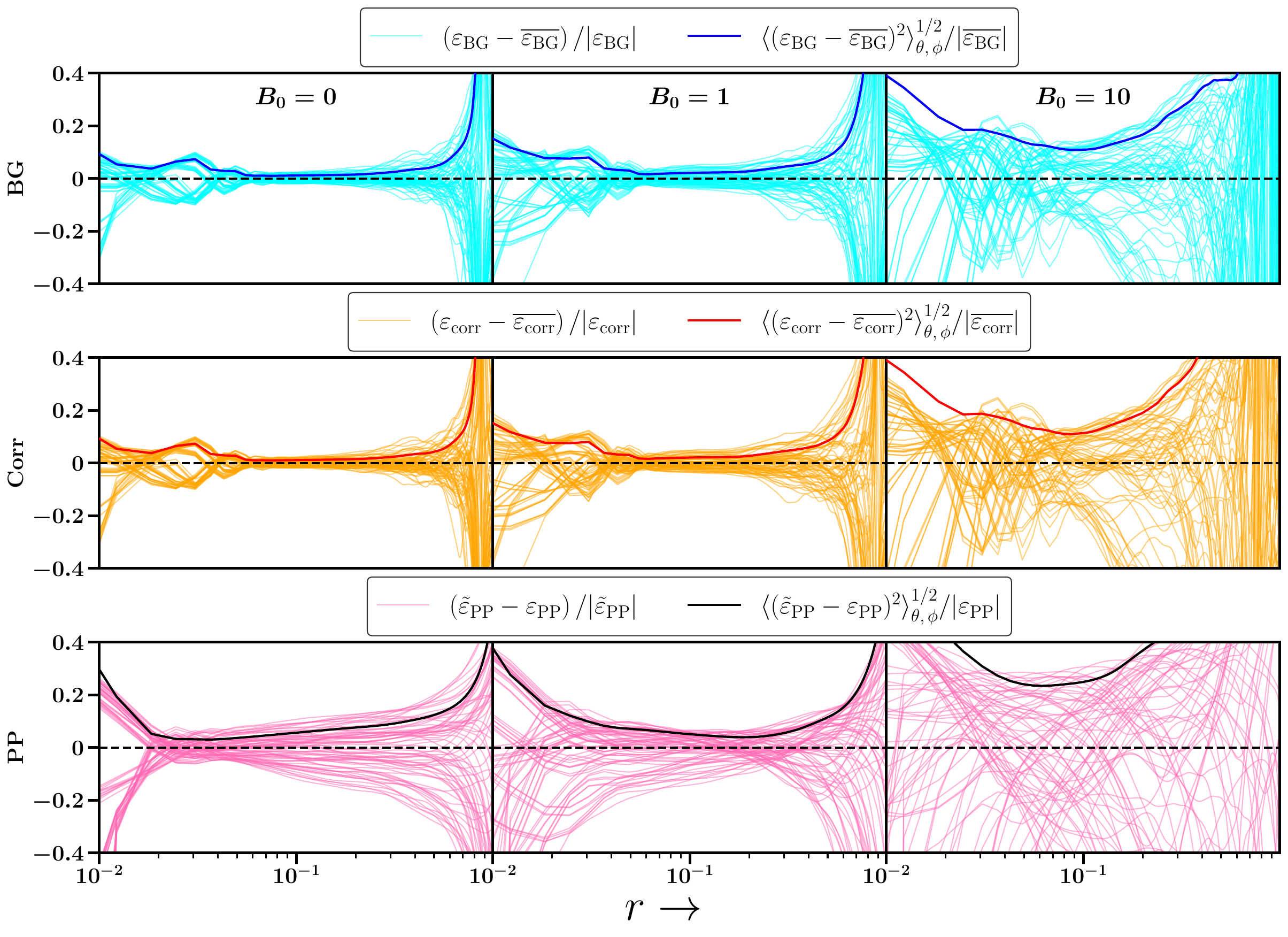}
    \caption{Normalized deviations of the directional cascade rates from their corresponding mean values (light-colored lines), together with the normalized root-mean-square angular deviations (dark-colored lines), for the BG (top), Corr (middle), and PP (bottom) formulations with different strengths of $B_0$.}
    \label{robust}
\end{figure*}

\subsection{Comparison between YM and PP formulations}
In this section, we compare the cascade rates obtained from the PP formulation with those obtained from the YM formulation. Now, integrating Eq. (\ref{divergence}) over a sphere of radius $r$, one obtains
\begin{align}
 &\int \varepsilon_\mathrm{YM} \,d\tau=\int \frac{1}{r^2}\frac{\partial}{\partial r}(r^2 F_r) \,d\tau\\
 &= \int^r_0\frac{\partial}{\partial r}\left(r^2\int_0^{2\pi}\int_0^\pi F_r\sin \theta\, d\theta d\phi \right)dr    = r^24\pi \langle F_r \rangle_{\theta, \,\phi}\nonumber,
\end{align}
\noindent where $d\tau$ is the elementary volume and
the contributions from the angular derivatives vanish upon integration over the entire solid angle as shown in Eqs. (\ref{theta}) and (\ref{phi}). If we further assume that the turbulent flow is statistically isotropic, $F_r \equiv F_r^{iso}$ does not depend on the angular coordinates $\theta$ and $\phi$ any longer and therefore $\langle F_r\rangle_{\theta,\phi}=F_r^{iso}$. A direct comparison with Eq. (\ref{politano}) clearly shows $\varepsilon_{\mathrm{PP}}= (3/r)F_r^{iso}=(3/4\pi r^3)\int \varepsilon_\mathrm{YM}\, d\tau$. For a flow that is not necessarily isotropic, such that $F_r^{\mathrm{iso}}\neq F_r$, one may nevertheless define an effective cascade rate, $\tilde{\varepsilon}_\mathrm{PP}(\bm{r}) = (3/r)F_r$. Although  $\tilde{\varepsilon}_\mathrm{PP}$ does not represent the true energy cascade rate obtained from the PP formulation under anisotropic conditions, it is analogous to the cascade-rate commonly obtained from single-spacecraft \textit{in situ} measurements and used to estimate the turbulent heating rate \citep{valvo_2007, MacBride_2008, Coburn_2014, Banerjee_2016}.

In Fig. \ref{pp_vs_ym}, we plot the pseudo-PP estimates $\tilde{\varepsilon}_\mathrm{PP}$ along the aforementioned 73 directions along with their angle-averaged value and $\overline{\varepsilon_\mathrm{YM}}$. For $B_0=0$, these directional estimates exhibit a weak spread which becomes more pronounced  as the strength of $B_0$ increases. Note that, unlike the BG and Corr formulations, which provide the complete nonlinear energy transfer rate for each separation vector $\bm{r}$, $\tilde{\varepsilon}_\mathrm{PP}$ does not provide a true directional transfer rate when the flow is not statistically isotropic. However, irrespective of the strengths of $B_0$, their angle-averaged value $\langle \tilde{\varepsilon}_\mathrm{PP} \rangle_{\theta, \,\phi}$ shows an excellent agreement with $\overline{\varepsilon_\mathrm{YM}}$ and  $\varepsilon_\mathrm{diss}$ in the inertial range, thereby providing a reliable estimate of the heating rate. This can be understood by noticing the fact that $\langle\tilde{\varepsilon}_{\mathrm{PP}}\rangle_{\theta,\phi}=(3/r)\langle F_r\rangle_{\theta,\, \phi}=(3/4\pi r^3)\int \varepsilon_\mathrm{YM}\, d\tau = \varepsilon_\mathrm{PP}$. It is important to note that one can still obtain a meaningful energy transfer rate from \textit{in situ} spacecraft measurements using the PP formulation if we have a large sample of data points spanning the full range of $\theta$ and $\phi$. Furthermore, by careful inspection, one observes an infrared shift of $\varepsilon_\mathrm{PP}$ with respect to the $\overline{\varepsilon_\mathrm{YM}}$  across the inertial scales. This behavior may arise from the integration of $\varepsilon_\mathrm{YM}$ to obtain $\varepsilon_\mathrm{PP}$, which acts as a low-pass filter incorporating small-scale contributions to the large scales.   
\begin{figure*}
    \centering
    \includegraphics[width=0.9
    \textwidth]{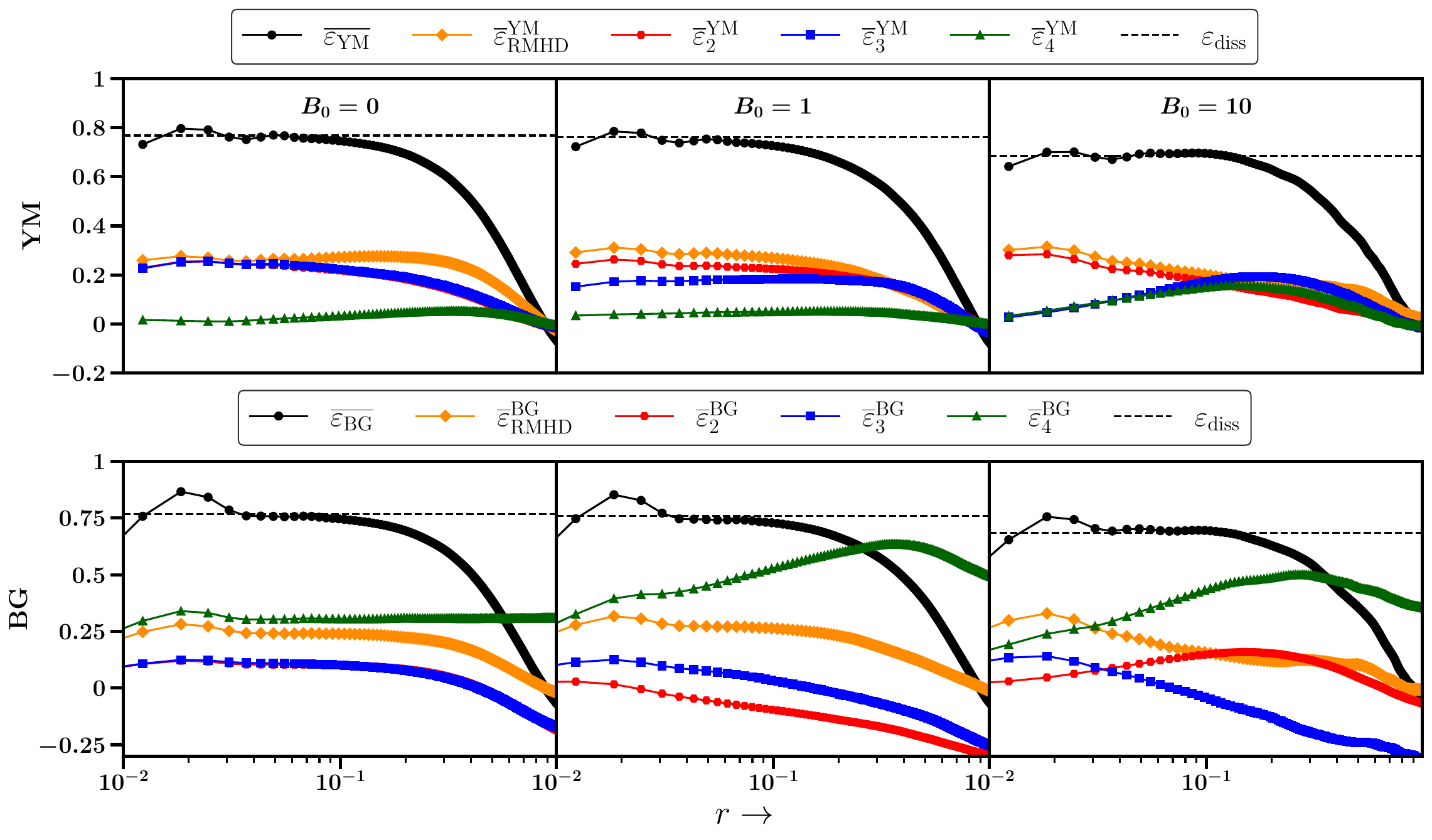}
    \caption{Decomposition of the YM and BG cascade rates into RMHD and non-RMHD contributions.
     Here, $\varepsilon_\mathrm{RMHD}^\mathrm{YM, \, BG}$ denotes the energy cascade rate in the RMHD framework, while ($\varepsilon_2^\mathrm{YM}, \, \varepsilon_3^\mathrm{YM}, \, \varepsilon_4^\mathrm{YM}$) and ($\varepsilon_2^\mathrm{BG}, \, \varepsilon_3^\mathrm{BG}, \, \varepsilon_4^\mathrm{BG}$) represent cascade rates, which involve parallel fluctuations as given in Eqs. (\ref{YM rmhd}) and (\ref{BG rmhd}), respectively.}
    \label{RMHD}
\end{figure*}

As discussed in the introduction, multi-spacecraft techniques enable the directional transfer rates to be evaluated using the BG and Corr formulations. In Fig. (\ref{robust}), we examine how these directional estimates deviate from their corresponding angle-averaged values and assess the extent to which they can serve as reliable proxies for the true energy cascade rate available for turbulent heating. In Fig. (\ref{robust}), we plot the normalized deviation of the directional transfer rates obtained from the BG and Corr formulations (upper and middle panels, respectively) from their corresponding angle-averaged values along with their root-mean-square deviations. Although \(\tilde{\varepsilon}_\mathrm{PP}\) does not represent the true directional energy transfer rate, we also show its deviation from true \(\varepsilon_\mathrm{PP}\), since \(\tilde{\varepsilon}_\mathrm{PP}\) has been widely employed in previous studies to estimate the turbulent energy transfer rate from \textit{in situ} spacecraft data. For $B_0=0$, the normalized deviations   $(\varepsilon_\mathrm{BG}-\overline{\varepsilon_\mathrm{BG}})/|\varepsilon_\mathrm{BG}|$ and $(\varepsilon_\mathrm{corr}-\overline{\varepsilon_\mathrm{corr}})/|\varepsilon_\mathrm{corr}|$ remain close to zero, indicating weak directional dependence. However, the deviations become non-negligible ($<5\%$ and $> 10\%$ for $B_0=1$ and 10, respectively) as the strength of $B_0$ increases, indicating a stronger directional dependence of the energy cascade rate. For PP, we also calculate $(\tilde{\varepsilon}_\mathrm{PP}-\varepsilon_\mathrm{PP})/|\tilde{\varepsilon}_\mathrm{PP}|$ and find a significant spread even in the absence of a mean magnetic field ($\sim5\%$ for $B_0=0$). The spread becomes increasingly pronounced with increasing $B_0$ ($\sim10\%$ and $> 25\%$ for $B_0=1$ and 10, respectively). However, note that unlike the BG and Corr formulations, here $(\tilde{\varepsilon}_\mathrm{PP}-\varepsilon_\mathrm{PP})/|\tilde{\varepsilon}_\mathrm{PP}|$ gives us a measure of deviation of the pseudo energy transfer rates with respect to the true $\varepsilon_\mathrm{PP}$. The result in Fig. \ref{robust} shows that in a plasma environment with moderate or strong mean magnetic field, the value of calculated $\tilde{\varepsilon}_\mathrm{PP}$ can deviate substantially from the true cascade rate $\varepsilon_\mathrm{PP}$.

\subsection{Contributions from RMHD vs. non-RMHD components}

As the strength of $B_0$ increases, the turbulence can effectively be described using the RMHD framework. However, previous studies have argued that, although a strong guide field is an essential element of conventional RMHD ordering, it does not, by itself, guarantee that the contributions of parallel fluctuations to the nonlinear energy transfer become negligible \citep{montogomery_1982, oughton_2017}. Recently, an exact relation for the inertial range energy cascade was derived for RMHD turbulence \citep{sasmal_2025}. In this paper, we therefore use the full-MHD simulations to directly quantify the fraction of the cascade carried by RMHD and non-RMHD components with increasing strength of $B_0$. To achieve this, we separate the RMHD and non-RMHD contributions to $\varepsilon_\mathrm{YM}$ and $\varepsilon_\mathrm{BG}$ as
\begin{align}
    \varepsilon_\mathrm{YM} &= \varepsilon_\mathrm{RMHD}^\mathrm{YM}+\varepsilon_2^\mathrm{YM}+\varepsilon_3^\mathrm{YM}+\varepsilon_4^\mathrm{YM},\label{YM rmhd}\text{ and }\\
    \varepsilon_\mathrm{BG}&=\varepsilon_\mathrm{RMHD}^\mathrm{BG}+\varepsilon_2^\mathrm{BG}+\varepsilon_3^\mathrm{BG}+\varepsilon_4^\mathrm{BG},\label{BG rmhd}
\end{align}
where 
\begin{align}
    &\varepsilon_\mathrm{RMHD}^{\mathrm{YM}}=\bnablar\cdot\left\langle2(\delta\bu_\perp\cdot\delta\bb_\perp)\delta\bb_\perp-\left[(\delta\bu_\perp)^2+(\delta\bb_\perp)^2\right]\delta\bu_\perp\right\rangle/4,\nonumber\\
    &\varepsilon_2^\mathrm{YM}=\bnablar\cdot\left\langle2(\delta\bu_\parallel\cdot\delta\bb_\parallel)\delta\bb_\perp-\left[(\delta\bu_\parallel)^2+(\delta\bb_\parallel)^2\right]\delta\bu_\perp\right\rangle/4,\nonumber\\
    &\varepsilon_3^\mathrm{YM}=\bnablar\cdot\left\langle2(\delta\bu_\perp\cdot\delta\bb_\perp)\delta\bb_\parallel-\left[(\delta\bu_\perp)^2+(\delta\bb_\perp)^2\right]\delta\bu_\parallel\right\rangle/4,\nonumber\\
&\varepsilon_4^\mathrm{YM}=\bnablar\cdot\left\langle2(\delta\bu_\parallel\cdot\delta\bb_\parallel)\delta\bb_\parallel-\left[(\delta\bu_\parallel)^2+(\delta\bb_\parallel)^2\right]\delta\bu_\parallel\right\rangle/4,\nonumber\\
&\varepsilon_\mathrm{RMHD}^\mathrm{BG}=\left\langle\delta\bu_\perp\cdot\delta(\bu_\perp\times\wz_\parallel+\jz_\parallel\times\bb_\perp)+\delta\jz_\parallel\cdot\delta(\bu_\perp\times\bb_\perp)\right\rangle/2,\nonumber\\
    &\varepsilon_2^\mathrm{BG}=\left\langle\delta\bu_\perp\cdot\delta(\bu_\parallel\times\wz_\perp+\jz_\perp\times\bb_\parallel)\right\rangle/2,\nonumber\\
    &\varepsilon_3^\mathrm{BG}=\left\langle\delta\bu_\parallel\cdot\delta(\bu_\perp\times\wz_\perp+\jz_\perp\times\bb_\perp)\right\rangle/2,\text{ and }\nonumber\\
    &\varepsilon_4^\mathrm{BG}=\left\langle\delta\jz_\perp\cdot\delta(\bu_\parallel\times\bb_\perp+\bu_\perp\times\bb_\parallel)\right\rangle/2,\nonumber
\end{align}
\noindent for any vector field $\bm{X}$, $\bm{X}_\parallel=(\bm{X}\cdot\hat{z})\hat{z}$ and $\bm{X}_\perp=\bm{X}-\bm{X}_\parallel$. Here, \(\varepsilon_{\rm RMHD}^{\rm YM,BG}\) denotes the contribution obtained by retaining the nonlinear terms consistent with the RMHD ordering. In Fig. (\ref{RMHD}), we plot the angle-averaged values for all these contributions corresponding to the three values of $B_0$. In the YM formulation, the results show $\overline{\varepsilon}_\mathrm{RMHD}^\mathrm{YM}$ is comparable to $\overline{\varepsilon}_2^\mathrm{YM}$ and in the inertial range, $\overline{\varepsilon}_\mathrm{RMHD}^\mathrm{YM}\approx\varepsilon_\mathrm{diss}/3$ for different strengths of $B_0$. Furthermore, ${\overline\varepsilon_3}^\mathrm{YM}$ becomes comparable to $\overline{\varepsilon}_\mathrm{RMHD}^\mathrm{YM}$ for $B_0=0$ and $1$, whereas for $B_0=10$, it closely follows $\overline{\varepsilon_4}^\mathrm{YM}$. In contrast, $\overline{\varepsilon_4}^\mathrm{YM}$ remains close to zero for the first two cases. In the BG formulation, $\overline{\varepsilon}_\mathrm{RMHD}^\mathrm{BG}$ also exhibits a behavior similar to the YM formulation. For $B_0=0$, the transfer rate associated with $\overline{\varepsilon_4}^\mathrm{BG}$ is comparable to $\overline{\varepsilon}_\mathrm{RMHD}^\mathrm{BG}$ in the inertial range, whereas in the cases with $B_0=1$ and $10$, it exceeds $\overline{\varepsilon}_\mathrm{RMHD}^\mathrm{BG}$. The results in Fig. (\ref{RMHD}) show that, even in the presence of a strong mean magnetic field, the contributions involving the parallel fluctuations are non-negligible. 
This is consistent with the fact that merely increasing the mean field from zero to a considerably strong magnitude is insufficient to recover the RMHD limit from incompressible MHD framework \citep{oughton_2017}.  

\begin{figure*}
    \centering
    \includegraphics[width=.9
    \textwidth]{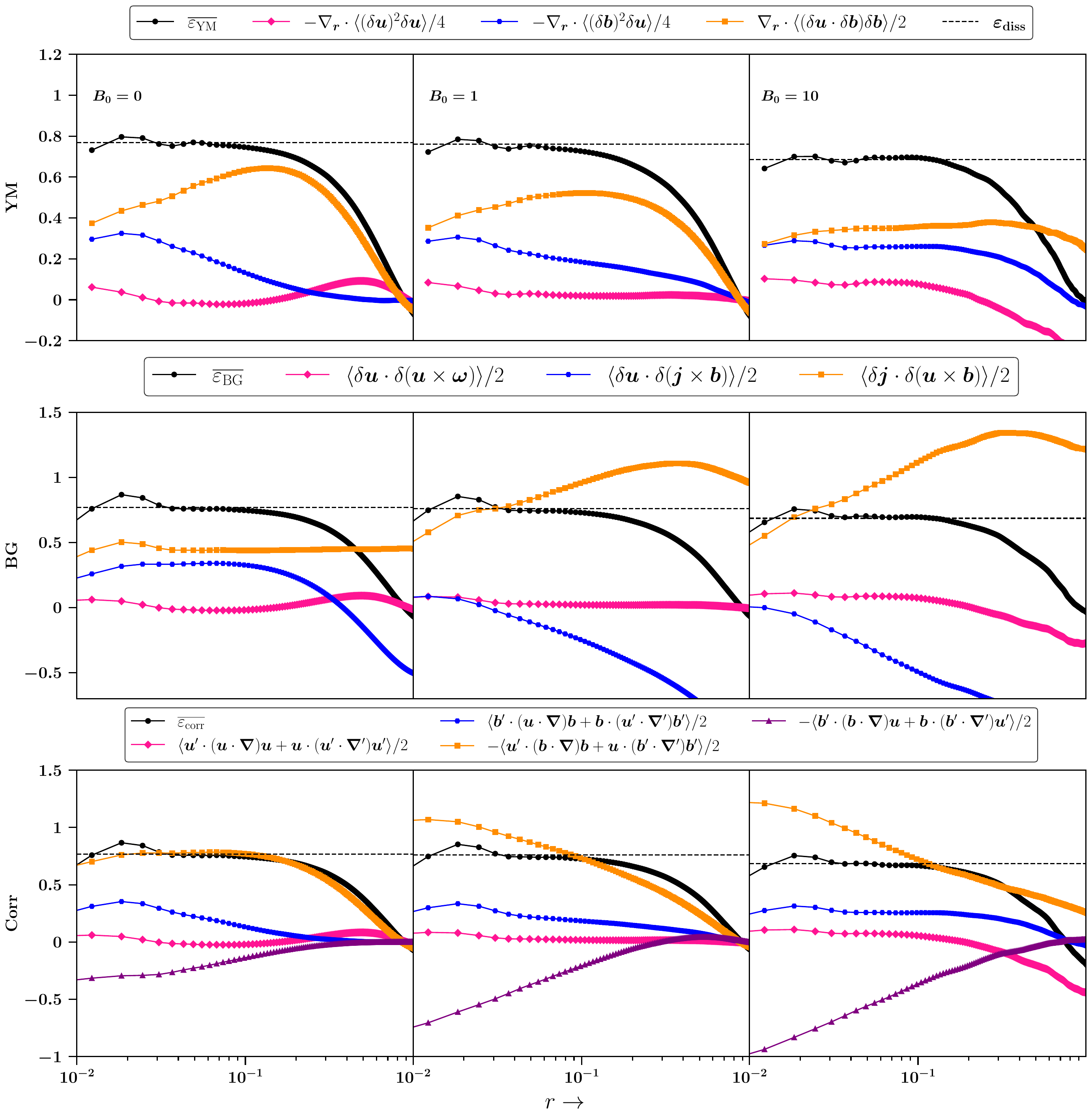}
    \caption{Term-by-term comparison of the contributions to the total energy cascade rate from different exact-law formulations, YM (top panels), BG (middle panels), and Corr (bottom panels), with $B_0=0, \, 1$, and $10$.}
    \label{term_comparison}
\end{figure*}

\subsection{Term-by-term analysis of different exact laws}
Proceeding one step further, we also study the individual contributions to the energy transfer rates (averaged over 73 directions) corresponding to different exact-law formulations (see Fig. \(\ref{term_comparison}\)). In the YM formulation, we observe that the term $\bnablar\cdot\left\langle(\delta\bu\cdot\delta\bb)\delta\bb\right\rangle/2$ (orange line) dominates the inertial range, although its contribution to the energy cascade decreases as the strength of $B_0$ increases. In contrast, the kinetic contribution to the total energy transfer rate, $-\bnablar\cdot\left\langle(\delta\bu)^2\delta\bu\right\rangle/4$ (pink line), is very small in the inertial range for different strengths of $B_0$. However, the results show the magnetic-advection contribution to the total transfer rate via the term $-\bnablar\cdot\left\langle(\delta\bb)^2\delta\bu\right\rangle/4$ (blue line) increases as energy cascades toward smaller scales for $B_0=0$, whereas it attains a constant value as $B_0$ increases. In the BG formulation, for $B_0=0$, both the transfer terms, $\left\langle\delta\bm{j}\cdot\delta(\bu\times\bb)\right\rangle/2$ (orange line) and $\left\langle\delta\bu\cdot\delta(\bm{j}\times\bb)\right\rangle/2$ (blue line), make significant contributions to the total energy cascade rate, whereas the kinetic energy transfer rate, $\left\langle\delta\bu\cdot\delta(\bu\times\bm{\omega})\right\rangle/2$ (pink line), remains close to zero for all three values of $B_0$. However, for $B_0=1\,\text{and } 10$, the transfer term corresponding to the orange line exceeds the total energy cascade rate in the inertial range, which is compensated by the reduction in the transfer rate associated with the blue line. In the Corr formulation, for the case with $B_0=0$, the transfer term, $-\left\langle\bu'\cdot(\bb\cdot\bnabla)\bb+\bu\cdot(\bb'\cdot\bnabla')\bb'\right\rangle/2$ (orange line), remains close to the total energy cascade rates, whereas for $B_0=1,\, \text{and } 10$, this transfer term surpasses the total energy transfer which is compensated by $-\left\langle\bb'\cdot(\bb\cdot\bnabla)\bu+\bb\cdot(\bb'\cdot\bnabla')\bu'\right\rangle/2$ (purple line). Note that the transfer terms associated with the orange and purple lines in the Corr formulation together yield the transfer rate represented by the orange line in the YM formulation. Here, the kinetic (pink line) and magnetic-advection (blue line) contributions to the total energy cascade rates, as expected, show behavior similar to that observed in the YM formulation.

\section{Summary and conclusion}\label{conclusion}

In this work, we have conducted the first-time numerical comparison of the three exact-law formulations of incompressible MHD turbulence, namely the YM, BG, and Corr formulations, with three strengths of the mean magnetic field $B_0$. Using high-resolution direct numerical simulations with $1024^3$ grid points, we have calculated the directional energy transfer rates $\varepsilon (r, \theta, \phi)$ and their angle-averaged values using both BG and Corr formulations. The observed directional spread is possibly due to the large scale contamination of the inertial range in the presence of a moderate or strong $B_0$. Despite the difficulty of calculating the directional transfer rates using YM formulation, we calculate the angle-averaged energy transfer rate from the YM formulation too and compare that with those obtained using the other two formulations. All of them show a scale-independent behaviour across the inertial scales where they become approximately equal to the average one-point dissipation rate $\varepsilon_{diss}$. In particular, for the globally non-stationary simulation with $B_0=10$, this behaviour hints at the possibility of quasi-stationarity at inertial scales. Importantly, an excellent agreement among the three angle-averaged cascade rates at all three strengths of $B_0$ presents the BG and Corr formulations as equally credible as the popular YM formulation while being particularly useful for calculating the turbulent heating rates in space and astrophysical plasmas involving anisotropy due to a non-negligible $B_0$. However, the use of BG and Corr formulations are subject to the calculation of the spatial gradients and therefore to the availability of multi-spacecraft measurements.  

A part of the current study is specially devoted to highlighting the caveat involved in the use of the isotropic PP formulation to estimate the energy cascade rate just from single-spacecraft measurements. Despite their frequent use in literature, we analytically argue that such an estimate does not represent a true directional energy transfer rate in the presence of a strong $B_0$, and deviates considerably from their angle-averaged value, which, nevertheless, provides a low-pass filtered estimate of angle-averaged cascade rate obtained from the YM formulation. 

By decomposing the YM and BG formulations into RMHD and non-RMHD contributions, we clearly show that the RMHD part contributes approximately to the one-third of the total energy cascade at all three strengths of the mean magnetic field. In fact, the contribution from $\left\langle\delta\jz_\perp\cdot\delta(\bu_\parallel\times\bb_\perp+\bu_\perp\times\bb_\parallel)\right\rangle/2$ is found to exceed the RMHD contribution across the inertial range. This supports the theoretical prediction that in incompressible MHD turbulence, the parallel fluctuations cannot be made negligible and hence the RMHD limit cannot be attained by merely increasing the strength of $B_0$ \citep{oughton_2017, sasmal_2025}. Finally, owing to a term-by-term analysis, we explore the contributions of different nonlinear interaction channels to the total energy transfer rate, which differ substantially from one another both in terms of magnitude and sign. In both YM and Corr formulations, the kinetic and magnetic-advection contributions to the total energy transfer separately attain scale-independent plateaux in the inertial range for $B_0=10$. In BG formulation, at zero mean magnetic field, each of the three contributions exhibit scale-independent behaviour in the inertial range. In all the three simulations, contributions involving the magnetic fields considerably dominate the pure kinetic contribution which is found to be comparatively weak. In case of moderate and strong $B_0$, the net scale-independent flux are found to emerge as a result of nontrivial balance between disparate transfer rates including some that exceeds the resultant transfer rates.   

The present study is important for future investigations in space plasma turbulence using the \textit{in situ} spacecraft measurements, for example, to compare the cascade rates obtained from different exact-law formulations in space plasmas with and without a mean magnetic field. One can also extend similar numerical studies for helical invariants in incompressible MHD turbulence. Upon the availability of multi-spacecraft data in near-sun solar wind in future, one can also estimate the cascade rate using the alternative formulations derived for finite-Larmor-radius MHD turbulence both numerically and observationally to obtain a reliable estimate of the rate of energy available for the selective ion heating in the solar wind \citep{sasmal_2026}.  

\section*{Acknowledgements}

We sincerely acknowledge the Smart-TURB project along with the developers and maintainers of the TURB-MHD database for making the high-resolution magnetohydrodynamic turbulence datasets publicly available and also for providing access to these valuable numerical datasets through the Smart-TURB portal. For data analysis, we acknowledge the support and resources provided by PARAM Sanganak under the National Supercomputing Mission of the 
Government of India, at the Indian Institute of Technology, Kanpur. SB acknowledges financial support from the Anusandhan National Research Foundation (ANRF), Government of India, through the Advanced Research Grant (ARG), Grant No. ANRF/PHY/2026106.

%\bibliography{citation}{}

\begin{thebibliography}{}
\expandafter\ifx\csname natexlab\endcsname\relax\def\natexlab#1{#1}\fi
\providecommand{\url}[1]{\href{#1}{#1}}
\providecommand{\dodoi}[1]{doi:~\href{http://doi.org/#1}{\nolinkurl{#1}}}
\providecommand{\doeprint}[1]{\href{http://ascl.net/#1}{\nolinkurl{http://ascl.net/#1}}}
\providecommand{\doarXiv}[1]{\href{https://arxiv.org/abs/#1}{\nolinkurl{https://arxiv.org/abs/#1}}}

% type= article
\bibitem[{R.~A. Antonia {et~al.}(1997)Antonia, Ould-Rouis, Anselmet, \& Zhu}]{Antonia_1997}
Antonia, R.~A., Ould-Rouis, M., Anselmet, F., \& Zhu, Y. 1997, \bibinfo{title}{Analogy between predictions of Kolmogorov and Yaglom,} Journal of Fluid Mechanics, 332, 395–409, \dodoi{10.1017/S0022112096004090}

% type= article
\bibitem[{P. Augier {et~al.}(2012)Augier, Galtier, \& Billant}]{augier_2012}
Augier, P., Galtier, S., \& Billant, P. 2012, \bibinfo{title}{Kolmogorov laws for stratified turbulence,} Journal of Fluid Mechanics, 709, 659–670, \dodoi{10.1017/jfm.2012.379}

% type= phdthesis
\bibitem[{S. Banerjee(2014)Banerjee}]{banerjee_thesis}
Banerjee, S. 2014, \bibinfo{title}{{Compressible turbulence in space and astrophysical plasmas : Analytical approach and in-situ data analysis for the solar wind},} Theses, {Universit{\'e} Paris Sud - Paris XI}, \dodoi{10.70675/c66e2a17zc154z4f9dzb87dz832ffe519953}

% type= article
\bibitem[{S. Banerjee {\&} S. Galtier(2013)Banerjee \& Galtier}]{banerjee_2013}
Banerjee, S., \& Galtier, S. 2013, \bibinfo{title}{Exact relation with two-point correlation functions and phenomenological approach for compressible magnetohydrodynamic turbulence,} Phys. Rev. E, 87, 013019, \dodoi{10.1103/PhysRevE.87.013019}

% type= article
\bibitem[{S. Banerjee {\&} S. Galtier(2014)Banerjee \& Galtier}]{banerjee_2014}
Banerjee, S., \& Galtier, S. 2014, \bibinfo{title}{A Kolmogorov-like exact relation for compressible polytropic turbulence,} Journal of Fluid Mechanics, 742, 230–242, \dodoi{10.1017/jfm.2013.657}

% type= article
\bibitem[{S. Banerjee {\&} S. Galtier(2016)Banerjee \& Galtier}]{banerjee_2017}
Banerjee, S., \& Galtier, S. 2016, \bibinfo{title}{An alternative formulation for exact scaling relations in hydrodynamic and magnetohydrodynamic turbulence,} Journal of Physics A: Mathematical and Theoretical, 50, 015501, \dodoi{10.1088/1751-8113/50/1/015501}

% type= article
\bibitem[{S. Banerjee {et~al.}(2016)Banerjee, Hadid, Sahraoui, \& Galtier}]{Banerjee_2016}
Banerjee, S., Hadid, L.~Z., Sahraoui, F., \& Galtier, S. 2016, \bibinfo{title}{SCALING OF COMPRESSIBLE MAGNETOHYDRODYNAMIC TURBULENCE IN THE FAST SOLAR WIND,} The Astrophysical Journal Letters, 829, L27, \dodoi{10.3847/2041-8205/829/2/L27}

% type= article
\bibitem[{S. Banerjee {et~al.}(2025)Banerjee, Halder, \& Das}]{halder_2025}
Banerjee, S., Halder, A., \& Das, A. 2025, \bibinfo{title}{Universal energy cascade and relaxation in three-dimensional inertial electron magnetohydrodynamic turbulence,} Physics of Plasmas, 32, 102307, \dodoi{10.1063/5.0280826}

% type= article
\bibitem[{S. Banerjee {et~al.}(2023)Banerjee, Halder, \& Pan}]{banerjee_2023}
Banerjee, S., Halder, A., \& Pan, N. 2023, \bibinfo{title}{Universal turbulent relaxation of fluids and plasmas by the principle of vanishing nonlinear transfers,} Phys. Rev. E, 107, L043201, \dodoi{10.1103/PhysRevE.107.L043201}

% type= article
\bibitem[{S. Banerjee {\&} A.~G. Kritsuk(2018)Banerjee \& Kritsuk}]{banerjee_2018}
Banerjee, S., \& Kritsuk, A.~G. 2018, \bibinfo{title}{Energy transfer in compressible magnetohydrodynamic turbulence for isothermal self-gravitating fluids,} Phys. Rev. E, 97, 023107, \dodoi{10.1103/PhysRevE.97.023107}

% type= misc
\bibitem[{D. Capocci {et~al.}(2025)Capocci, Biferale, Bonaccorso, \& Linkmann}]{capocci_2025}
Capocci, D., Biferale, L., Bonaccorso, F., \& Linkmann, M. 2025, TURB-MHD: an open-access database of forced homogeneous magnetohydrodynamic turbulence, \doarXiv{2504.10755}

% type= article
\bibitem[{A. Celani {et~al.}(1998{\natexlab{a}})Celani, Prandi, \& Boffetta}]{celani_1998a}
Celani, A., Prandi, R., \& Boffetta, G. 1998{\natexlab{a}}, \bibinfo{title}{Kolmogorov's law for two-dimensional electron-magnetohydrodynamic turbulence,} Europhysics Letters, 41, 13, \dodoi{10.1209/epl/i1998-00109-0}

% type= article
\bibitem[{A. Celani {et~al.}(1998{\natexlab{b}})Celani, Prandi, \& Boffetta}]{celani_1998b}
Celani, A., Prandi, R., \& Boffetta, G. 1998{\natexlab{b}}, \bibinfo{title}{Turbulence in two-dimensional forced electron-magnetohydrodynamics,} Physica Scripta, 1998, 191, \dodoi{10.1238/Physica.Topical.075a00191}

% type= article
\bibitem[{J.~T. Coburn {et~al.}(2014)Coburn, Smith, Vasquez, Forman, \& Stawarz}]{Coburn_2014}
Coburn, J.~T., Smith, C.~W., Vasquez, B.~J., Forman, M.~A., \& Stawarz, J.~E. 2014, \bibinfo{title}{VARIABLE CASCADE DYNAMICS AND INTERMITTENCY IN THE SOLAR WIND AT 1 AU,} The Astrophysical Journal, 786, 52, \dodoi{10.1088/0004-637X/786/1/52}

% type= article
\bibitem[{T. {de Karman} {\&} L. {Howarth}(1938){de Karman} \& {Howarth}}]{karman_1938}
{de Karman}, T., \& {Howarth}, L. 1938, \bibinfo{title}{{On the Statistical Theory of Isotropic Turbulence},} Proceedings of the Royal Society of London Series A, 164, 192, \dodoi{10.1098/rspa.1938.0013}

% type= article
\bibitem[{D. Deepali {\&} S. Banerjee(2021)Deepali \& Banerjee}]{deepali_2021}
Deepali, D., \& Banerjee, S. 2021, \bibinfo{title}{Scale-dependent anisotropy of electric field fluctuations in solar wind turbulence,} Monthly Notices of the Royal Astronomical Society: Letters, 504, L1, \dodoi{10.1093/mnrasl/slab027}

% type= article
\bibitem[{M. Dunlop {et~al.}(1988)Dunlop, Southwood, Glassmeier, \& Neubauer}]{dunlop_1988}
Dunlop, M., Southwood, D., Glassmeier, K.-H., \& Neubauer, F. 1988, \bibinfo{title}{Analysis of multipoint magnetometer data,} Advances in Space Research, 8, 273, \dodoi{https://doi.org/10.1016/0273-1177(88)90141-X}

% type= article
\bibitem[{M.~W. Dunlop {et~al.}(2015)Dunlop, Yang, Yang, Xiong, Lühr, Bogdanova, Shen, Olsen, Zhang, Cao, Fu, Liu, Carr, Ritter, Masson, \& Haagmans}]{dunlop_2015}
Dunlop, M.~W., Yang, J.-Y., Yang, Y.-Y., {et~al.} 2015, \bibinfo{title}{Simultaneous field-aligned currents at Swarm and Cluster satellites,} Geophysical Research Letters, 42, 3683, \dodoi{https://doi.org/10.1002/2015GL063738}

% type= phdthesis
\bibitem[{R. Ferrand(2021)Ferrand}]{ferrand_2021}
Ferrand, R. 2021, \bibinfo{title}{{Multi-scale compressible turbulence in astrophysical plasmas viewed through theoretical, numerical and observational methods},} Theses, {Universit{\'e} Paris-Saclay}, \dodoi{10.70675/98611f2ez6c86z4409zb200z4f20e4852967}

% type= article
\bibitem[{R. Ferrand {et~al.}(2019)Ferrand, Galtier, Sahraoui, Meyrand, Andrés, \& Banerjee}]{ferrand_2019}
Ferrand, R., Galtier, S., Sahraoui, F., {et~al.} 2019, \bibinfo{title}{On Exact Laws in Incompressible Hall Magnetohydrodynamic Turbulence,} The Astrophysical Journal, 881, 50, \dodoi{10.3847/1538-4357/ab2be9}

% type= book
\bibitem[{U. Frisch(1995)Frisch}]{Frisch_1995}
Frisch, U. 1995, Turbulence: The Legacy of A. N. Kolmogorov (Cambridge University Press)

% type= article
\bibitem[{S. Galtier {\&} S. Banerjee(2011)Galtier \& Banerjee}]{banerjee_2011}
Galtier, S., \& Banerjee, S. 2011, \bibinfo{title}{Exact Relation for Correlation Functions in Compressible Isothermal Turbulence,} Phys. Rev. Lett., 107, 134501, \dodoi{10.1103/PhysRevLett.107.134501}

% type= article
\bibitem[{T.~S. Horbury {et~al.}(2008)Horbury, Forman, \& Oughton}]{horbury_2008}
Horbury, T.~S., Forman, M., \& Oughton, S. 2008, \bibinfo{title}{Anisotropic Scaling of Magnetohydrodynamic Turbulence,} Phys. Rev. Lett., 101, 175005, \dodoi{10.1103/PhysRevLett.101.175005}

% type= article
\bibitem[{A. Kolmogorov(1991)Kolmogorov}]{kolmogorov_1941}
Kolmogorov, A. 1991, \bibinfo{title}{Dissipation of energy in the locally isotropic turbulence,} Proceedings of the Royal Society of London. Series A: Mathematical and Physical Sciences, 434, 15, \dodoi{10.1098/rspa.1991.0076}

% type= article
\bibitem[{B.~T. MacBride {et~al.}(2008)MacBride, Smith, \& Forman}]{MacBride_2008}
MacBride, B.~T., Smith, C.~W., \& Forman, M.~A. 2008, \bibinfo{title}{The Turbulent Cascade at 1 AU: Energy Transfer and the Third-Order Scaling for MHD,} The Astrophysical Journal, 679, 1644, \dodoi{10.1086/529575}

% type= article
\bibitem[{W.~H. Matthaeus {et~al.}(1996{\natexlab{a}})Matthaeus, Bieber, \& Zank}]{matthaeus_1996}
Matthaeus, W.~H., Bieber, J.~W., \& Zank, G.~P. 1996{\natexlab{a}}, \bibinfo{title}{Anisotropic turbulence in the solar wind,} AIP Conference Proceedings, 382, 256, \dodoi{10.1063/1.51394}

% type= article
\bibitem[{D. {Montgomery}(1982){Montgomery}}]{montogomery_1982}
{Montgomery}, D. 1982, \bibinfo{title}{{Major Disruptions, Inverse Cascades, and the Strauss Equations},} Physica Scripta Volume T, 2A, 83, \dodoi{10.1088/0031-8949/1982/T2A/009}

% type= article
\bibitem[{S. Mouraya {\&} S. Banerjee(2019)Mouraya \& Banerjee}]{mouraya_2019}
Mouraya, S., \& Banerjee, S. 2019, \bibinfo{title}{Determination of energy flux rate in homogeneous ferrohydrodynamic turbulence using two-point statistics,} Phys. Rev. E, 100, 053105, \dodoi{10.1103/PhysRevE.100.053105}

% type= article
\bibitem[{K.~T. Osman {et~al.}(2011)Osman, Wan, Matthaeus, Weygand, \& Dasso}]{osman_2011}
Osman, K.~T., Wan, M., Matthaeus, W.~H., Weygand, J.~M., \& Dasso, S. 2011, \bibinfo{title}{Anisotropic Third-Moment Estimates of the Energy Cascade in Solar Wind Turbulence Using Multispacecraft Data,} Phys. Rev. Lett., 107, 165001, \dodoi{10.1103/PhysRevLett.107.165001}

% type= article
\bibitem[{S. Oughton {et~al.}(2017)Oughton, Matthaeus, \& Dmitruk}]{oughton_2017}
Oughton, S., Matthaeus, W.~H., \& Dmitruk, P. 2017, \bibinfo{title}{Reduced MHD in Astrophysical Applications: Two-dimensional or Three-dimensional?} The Astrophysical Journal, 839, 2, \dodoi{10.3847/1538-4357/aa67e2}

% type= article
\bibitem[{N. Pan {\&} S. Banerjee(2022)Pan \& Banerjee}]{pan_2022}
Pan, N., \& Banerjee, S. 2022, \bibinfo{title}{Exact relations for energy transfer in simple and active binary fluid turbulence,} Phys. Rev. E, 106, 025104, \dodoi{10.1103/PhysRevE.106.025104}

% type= article
\bibitem[{N. Pan {\&} S. Banerjee(2025)Pan \& Banerjee}]{pan_2025}
Pan, N., \& Banerjee, S. 2025, \bibinfo{title}{Universal energy cascade in homogeneous binary fluid turbulence: A direct comparison of different exact relations,} Phys. Rev. Fluids, 10, 064615, \dodoi{10.1103/pfwf-cp3p}

% type= article
\bibitem[{F. Pecora {et~al.}(2023)Pecora, Yang, Matthaeus, Chasapis, Klein, Stevens, Servidio, Greco, Gershman, Giles, \& Burch}]{Pecora_2023}
Pecora, F., Yang, Y., Matthaeus, W.~H., {et~al.} 2023, \bibinfo{title}{Three-Dimensional Energy Transfer in Space Plasma Turbulence from Multipoint Measurement,} Phys. Rev. Lett., 131, 225201, \dodoi{10.1103/PhysRevLett.131.225201}

% type= article
\bibitem[{J.~J. Podesta(2008)Podesta}]{podesta_2008}
Podesta, J.~J. 2008, \bibinfo{title}{Laws for third-order moments in homogeneous anisotropic incompressible magnetohydrodynamic turbulence,} Journal of Fluid Mechanics, 609, 171–194, \dodoi{10.1017/S0022112008002280}

% type= article
\bibitem[{H. Politano {\&} A. Pouquet(1998{\natexlab{a}})Politano \& Pouquet}]{politano_1998b}
Politano, H., \& Pouquet, A. 1998{\natexlab{a}}, \bibinfo{title}{Dynamical length scales for turbulent magnetized flows,} Geophysical Research Letters, 25, 273, \dodoi{https://doi.org/10.1029/97GL03642}

% type= article
\bibitem[{H. Politano {\&} A. Pouquet(1998{\natexlab{b}})Politano \& Pouquet}]{politano_1998a}
Politano, H., \& Pouquet, A. 1998{\natexlab{b}}, \bibinfo{title}{von K\'arm\'an--Howarth equation for magnetohydrodynamics and its consequences on third-order longitudinal structure and correlation functions,} Phys. Rev. E, 57, R21(R), \dodoi{10.1103/PhysRevE.57.R21}

% type= book
\bibitem[{S.~B. Pope(2000)Pope}]{Pope_2000}
Pope, S.~B. 2000, Turbulent Flows (Cambridge University Press)

% type= article
\bibitem[{S. Sakshee {et~al.}(2022)Sakshee, Bandyopadhyay, \& Banerjee}]{sakshee_2022}
Sakshee, S., Bandyopadhyay, R., \& Banerjee, S. 2022, \bibinfo{title}{MHD-scale anisotropy in solar wind turbulence near the Sun using Parker solar probe data,} Monthly Notices of the Royal Astronomical Society, 514, 1282, \dodoi{10.1093/mnras/stac1449}

% type= article
\bibitem[{R. Sasmal {\&} S. Banerjee(2025)Sasmal \& Banerjee}]{sasmal_2025}
Sasmal, R., \& Banerjee, S. 2025, \bibinfo{title}{Universal cascade and relaxation of strong anisotropic turbulence in plasma fluids,} Physics of Fluids, 37, 025197, \dodoi{10.1063/5.0250255}

% type= misc
\bibitem[{R. Sasmal {\&} S. Banerjee(2026)Sasmal \& Banerjee}]{sasmal_2026}
Sasmal, R., \& Banerjee, S. 2026, Determination of turbulent heating rate and relaxed states in finite Larmor radius magnetohydrodynamic turbulence with helicity barrier, \doarXiv{2604.28165}

% type= article
\bibitem[{R. {Singh} {et~al.}(2026){Singh}, {Banerjee}, \& {Halder}}]{singh_2026}
{Singh}, R., {Banerjee}, S., \& {Halder}, A. 2026, \bibinfo{title}{{An integration-free method for calculating curl and divergence in space plasmas using multi-spacecraft data},} \aap, 712, A134, \dodoi{10.1051/0004-6361/202558648}

% type= misc
\bibitem[{{}{Smart-Turb}(2020){Smart-Turb}}]{smart_turb}
{Smart-Turb}. 2020, \url{https://smart-turb.roma2.infn.it/init/routes/#/logging/welcome}

% type= article
\bibitem[{L. Sorriso-Valvo {et~al.}(2007)Sorriso-Valvo, Marino, Carbone, Noullez, Lepreti, Veltri, Bruno, Bavassano, \& Pietropaolo}]{valvo_2007}
Sorriso-Valvo, L., Marino, R., Carbone, V., {et~al.} 2007, \bibinfo{title}{Observation of Inertial Energy Cascade in Interplanetary Space Plasma,} Phys. Rev. Lett., 99, 115001, \dodoi{10.1103/PhysRevLett.99.115001}

% type= article
\bibitem[{M.~A. Taylor {et~al.}(2003)Taylor, Kurien, \& Eyink}]{taylor_2003}
Taylor, M.~A., Kurien, S., \& Eyink, G.~L. 2003, \bibinfo{title}{Recovering isotropic statistics in turbulence simulations: The Kolmogorov 4/5th law,} Phys. Rev. E, 68, 026310, \dodoi{10.1103/PhysRevE.68.026310}

\end{thebibliography}
\bibliographystyle{aasjournalv7.1}

\end{document}